\documentclass[3p,times,twocolumn]{elsarticle}
\usepackage{ecrc-mod}
\biboptions{sort,compress}

\volume{(to be published in Nucl. Phys. B Proc. Suppl.)}
\firstpage{1}
\journalname{HEP-version}

\runauth{Alexander\,P.\,Bakulev and Irina\,V.\,Potapova}

\usepackage{amssymb,amsfonts,amsmath,bm}
\usepackage{color}

\graphicspath{{./figs-bakulev/}}
\newcommand{\Gree}[1]{\textcolor[cmyk]{0.76,0,0.76,0.45}{#1}}
\newcommand{\Ds}{\displaystyle}

\begin{document}
\begin{frontmatter}
\dochead{}
\title{Resummation Approach in QCD Analytic Perturbation Theory}

\author{Alexander\,P.\,Bakulev}

\author{Irina\,V.\,Potapova\corref{cor2}}
 \cortext[cor2]{Responsible for Sect.~\ref{sec:Resum.FAPT.3L} and \ref{sec:Resum.Adler}}
  \address{Bogoliubov Laboratory of Theoretical Physics, JINR (Dubna, Russia)}

\begin{abstract}
 We discuss the resummation approach
 in QCD Analytic Perturbation Theory (APT).
 We start with a simple example of asymptotic power series
 for a zero-dimensional analog of the scalar $g\,\varphi^4$ model.
 Then we give a short historic preamble of APT
 and show that renormgroup improvement
 of the QCD perturbation theory
 dictates to use the Fractional APT (FAPT).
 After that we discuss the (F)APT resummation of nonpower series
 and provide the one-, two-, and three-loop
 resummation recipes.
 We show the results of applications of these recipes
 to the estimation of the Adler function $D(Q^2)$
 in the $N_f=4$ region of $Q^2$
 and
 of the Higgs-boson-decay width $\Gamma_{H\to b\overline{b}}(m_H^2)$
 for $M_H=100-180$~GeV$^2$.
\end{abstract}

\end{frontmatter}

\section{Simple example of asymptotic power series}
 \label{sec:fi4}
  In spite of many examples of successful applications
of perturbative approach in quantum field theory,
a perturbative power expansion for a quantum amplitude
usually
is not convergent.
Instead, a typical power series
appears to be asymptotic.
To refresh the reader knowledge on this subject
we consider a simple example
--- the so-called  ``0-dimensional'' analog
of the scalar field theory
$g \varphi^4$:
\begin{subequations}
\begin{eqnarray}
 \label{eq:int.0D}
  I(g)=\int_{\!\!-\infty}^{\,\,\infty} e^{-x^2-gx^4}\,d x\,.
\end{eqnarray}
It can be expanded in a power series~\cite{KaSh80}
\begin{eqnarray}
 \label{eq:I-series}
   I(g) = \sum_{k=0}(-g)^k I_k\,,
\end{eqnarray}
with factorially growing coefficients:
$I_k = \Gamma(2k+1/2)/\Gamma(k+1)\,\to\, 2^k\,k!$
for $k\gg1$.
Meanwhile, $I(g)$
can be expressed via special MacDonald function
\begin{eqnarray}
 \label{eq:I-McDon}
 I(g) =
  \frac{1}{2\,\sqrt{g}}\,e^{1/8g}\,
 K_{1/4}\left(\frac{1}{8g}\right)
\end{eqnarray}
with known analytic properties in the complex $g$ plane:
It is a four-sheeted function  analytical
in the whole complex plane
besides the cut
from the origin $g=0$ along the whole negative semiaxis.
It has an essential singularity
$\displaystyle{e^{-1/8g}}$
at the origin
and in its vicinity on the first Riemann sheet
it can be written down
in the Cauchy integral form:
\begin{eqnarray}
 \label{eq:I-CInt}
  I(g) =
   \sqrt{\pi}
   - \frac{g}{\sqrt{2\,\pi}}\!
      \int_0^{\infty}
       \frac{d\gamma\,e^{-1/4\gamma}}{\gamma(g+\gamma)}\,.
\end{eqnarray}
\end{subequations}
Due to this singular behavior near the origin
the power Taylor series (\ref{eq:I-series})
has no convergence domain
for real positive $g$ values.
This behavior is in one-to-one correspondence with
the factorial growth of power expansion coefficients.
The same factorial growth of expansion coefficients
has been proved
for the $\phi^4$ scalar and a few other QFT models~\cite{Lip76}.

\begin{table*}[!t]
  \centering
 \caption{The last detained ($(-g)^{K}\,I_{K}$) and the first
  dismissed ($(-g)^{K+1}\,I_{K+1}$) contributions to the power series (\ref{eq:I-series})
  in comparison with the exact value (\ref{eq:I-McDon}) and the approximate result $I_K(g)$,
  with $K$ being the truncation number and $\Delta_K I(g)$ --- the error of approximation.
 \label{tab:I(g)}\vspace*{+2.3mm}}
\begin{tabular}{|c||c||c|c||c|c|c|}\hline
$g\vphantom{^|_|}$
      &$K$
             &$(-g)^{K}\,I_{K}$
                    &$(-g)^{K+1}\,I_{K+1}$
                           &$I_{K}(g)$
                                  &$I(g)$
                                         &$\Delta_K I(g)$\\ \hline\hline
$0.07$
      &\Gree{$\bm{7}$}
             &$-0.04 (2\%)$
                    &$+0.07 (4.4\%)\vphantom{^|_|}$
                           &$1.674$
                                  &$1.698$
                                         &\Gree{$\bm{1.4\%}$} \\ \hline
$0.07$
      &$9$
             &$-0.17 (10\%)$
                    &$+0.42 (25\%)\vphantom{^|_|}$
                           &$1.582$
                                  &$1.698$
                                         &$7\%$ \\ \hline
$0.15$
      &\Gree{$\bm{2}$}
             &$+0.13 (8\%)$
                    &$-0.16 (10\%)\vphantom{^|_|}$
                           &$1.704$
                                  &$1.639$
                                         &\Gree{$\bm{4\%}$} \\ \hline
$0.15$
      &$4$
             &$+0.30 (18\%)$
                    &$-0.72 (44\%)\vphantom{^|_|}$
                           &$1.838$
                                  &$1.639$
                                         &$12\%$ \\ \hline
 \end{tabular}
\end{table*}

Power series with
factorially growing coefficients
belongs to the class of Asymptotic Seria (AS) ---
their properties were investigated
by
Henry Poincar\'e at the end of the XIX century.
In short,
he concluded
that the truncated AS
can be used for obtaining the quantitative information
on expanded function.
To be more concrete,
the error of approximating $F(g)$
by first $K$ terms of expansion,
$F(g)\to F_K(g)=\Sigma_{k\leq K}\,f_k(g)$,
is equal to the last accounted term $f_K(g)$.
This observation can be used
to obtain a lower limit of possible accuracy
for the given $g$ value:
one should find the number $K$
with the minimal value of $f_K(g)$ ---
then its absolute value just
set the limit of accuracy.

To make this statement more clear,
we take a power AS,
$f_k(g)=f_k\,g^k$,
with factorial growth of expansion coefficients,
$f_k\sim k!$.
Then we see that terms $f_k(g)$ stop to decay
at $k \sim K\sim 1/g$,
hence
for $k\gtrsim K+1$ truncation error starts to grow.
The explicit calculations for the function $I(g)$
(\ref{eq:int.0D})
with AS (\ref{eq:I-series}),
presented in Table~\ref{tab:I(g)} (see also in~\cite{BaSh11}),
shows
that the optimal values of truncation number
$K=K_*\simeq1/(2g)$ (in dark green)
indeed provides the best possible accuracy
(dark green in the last column):
Attempting to account a couple of extra terms
(the second and the forth lines)
results in drastic rise of the error!
Thus, one has $\,K_*(g=0.07)=7\,$
and $\,K_*(g=0.15)=2\,$.
It is not possible at all to get the 1\% accuracy
for $g=0.15\,$,
instead the best accuracy here is 4\%.
In this situation
to obtain more accurate result
we need to use Eq.\,(\ref{eq:I-McDon})
which has the right analytic properties in $g$.

We conclude from this example
that to produce reliable predictions for a physical amplitude
in QCD
one needs a tool for resumming the original perturbative power series
into the quasi-nonperturbative amplitude
which should have prescribed analytical properties,
dictated by the first principles of QCD,
like causality and renormalizability.
It is intriguing
that the APT approach in QCD allows one
to obtain such a tool.

\section{Analytic Perturbation Theory in QCD}
\label{sec:APT}
In the standard QCD PT
we know the Renormalization Group (RG) equation
in the $l$-loop approximation
\begin{eqnarray}
 \label{eq:RG.a}
  \frac{d a_{(l)}[L]}{d L}
   = - a_{(l)}^2[L]\left[1+\sum_{k\geq1}^{l-1}c_k\,a_{(l)}^k[L]\right]
\end{eqnarray}
for the effective coupling $\alpha_s(Q^2)=a_{(l)}[L]/\beta_f$
with $L=\ln(Q^2/\Lambda^2)$, $\beta_f=b_0(N_f)/(4\pi)=(11-2N_f/3)/(4\pi)$.\footnote{%
   We use notations $f(Q^2)$ and $f[L]$ in order to specify the arguments we mean ---
   squared momentum $Q^2$ or its logarithm $L=\ln(Q^2/\Lambda^2)$,
   that is $f[L]=f(\Lambda^2\cdot e^L)$ and the QCD scale parameter $\Lambda$
   is usually referred to $N_f=3$ region.}
Then its one-loop solution generates Landau pole singularity,
$a_{(1)}[L] = 1/L$.
As a consequence,
the perturbative power series for the Adler function,
$D(Q^2)=d_0+\sum_{k\geq1}d_k\,a^k[L]$,
being reasonably good in the deep UV region of $Q^2$,
at $Q^2=\Lambda^2$ has the unphysical pole
--- in marked contrast to the analyticity property
of the Adler function.
Indeed, it should have a cut along the negative axis of $Q^2$,
but can not have any poles in Euclidean domain, $Q^2>0$.
APT was suggested as a resolution of this contradiction:
From the very beginning we demand
that improved coupling,
as well as all its integer powers,
should have the right analytic properties,
i.\,e.
be represented as dispersive integrals,
see (\ref{eq:An.SD}).

Strictly speaking the QCD Analytic Perturbation Theory (APT)
was initiated by N.~N.~Bogoliubov et al. paper of 1959~\cite{BLS60},
where ghost-free effective coupling for QED has been constructed.
Then in 1982 Radyushkin \cite{Rad82}
and Krasnikov and Pivovarov \cite{KP82}
using the same dispersion technique
suggested regular (for $s\geq \Lambda^2$) QCD running coupling
in Minkowskian region,
the well-known $\pi^{-1}\arctan(\pi/L)$.
After that in 1995 Jones and Solovtsov using variational approach~\cite{JS95-349}
constructed the effective couplings in Euclidean and Minkowski domains
which appears to be finite for all $Q^2$ and $s$
and satisfy analyticity integral conditions (\ref{eq:R-D-operation}).
Just in the same time Shirkov and Solovtsov \cite{SS},
using the dispersion approach of~\cite{BLS60},
discovered ghost-free coupling $\mathcal A_1[L]$, Eq.\ (\ref{eq:A_1}),
in Euclidean region
and
ghost-free coupling $\mathfrak A_1[L]$, Eq.\ (\ref{eq:U_1}),
in Minkowskian region,
which satisfy analyticity integral conditions
\begin{subequations}
 \label{eq:R-D-operation}
\begin{eqnarray}
 \label{eq:D-operation}
 \!\!\!\!\!\!\!\!\!\mathcal A_1(Q^2)
  = \hat{D}\left[\mathfrak A_1\right](Q^2)
  \equiv
   Q^2~\int_0^{\infty}
    \frac{\mathfrak A_1(\sigma)}{(\sigma+Q^2)^2}
     d\sigma;\!\!\!\\
 \label{eq:R-operation}
 \!\!\!\!\!\!\!\!\!\mathfrak A_1(s)
  = \hat{R}\left[\mathcal A_1\right](s)
  \equiv
   \frac{1}{2\pi i}
    \int_{-s-i\varepsilon}^{-s+i\varepsilon}\!
     \frac{\mathcal A_1(\sigma)}{\sigma}\,
  d\sigma.\!\!\!
\end{eqnarray}
\end{subequations}
The last coupling coincides with the Radyushkin one for $s\geq\Lambda^2$.
Due to the absence of singularities in these couplings,
Shirkov and Solovtsov suggested to use them for all $Q^2$ and $s$
(for recent applications --- see in~\cite{PST}).

Shirkov--Solovtsov approach, now termed APT,
appears to be very powerful:
in Euclidean domain,
$\displaystyle-q^2=Q^2$, $\displaystyle L=\ln Q^2/\Lambda^2$,
it generates the following set of images for the effective coupling
and its $n$-th powers,
$\displaystyle\left\{{\mathcal A}_n[L]\right\}_{n\in\mathbb{N}}$,
whereas in Minkowskian domain,
$\displaystyle q^2=s$, $\displaystyle L_s=\ln s/\Lambda^2$,
it generates another set,
$\displaystyle\left\{{\mathfrak A}_n[L_s]\right\}_{n\in\mathbb{N}}$.
APT is based on the RG and causality
that guaranties standard perturbative UV asymptotics
and spectral properties.
Power series of the standard PT
$\sum_{m}d_m a^m[L]$
transforms into non-power series
$\sum_{m}d_m {\mathcal A}_{m}[L]$ in APT.

By the analytization in APT for an observable $f(Q^2)$
we mean the dispersive ``K\"allen--Lehman'' representation
\begin{eqnarray}
 \label{eq:An.SD}
  \left[f(Q^2)\right]_\text{an}
   = \int_0^{\infty}\!
      \frac{\rho_f(\sigma)}
         {\sigma+Q^2-i\epsilon}\,
       d\sigma
\end{eqnarray}
with $\Ds\rho_f(\sigma)={\pi}^{-1}\,\textbf{Im}\,\big[f(-\sigma)\big]$.
Then in the one-loop approximation $\rho_1^{(1)}(\sigma)=1/\sqrt{L_\sigma^2+\pi^2}$
and
\begin{subequations}
 \label{eq:A.U}
 \begin{eqnarray}
  \label{eq:A_1}
   \!\!\!\!\!\!\!\!\!\mathcal A_1^{(1)}[L]
   \!\!&\!\!=\!\!&\!\! \int_0^{\infty}\!\frac{\rho_1(\sigma)}{\sigma+Q^2}\,d\sigma
            = \frac{1}{L} - \frac{1}{e^L-1},\!\!\!\\
  \label{eq:U_1}
   \!\!\!\!\!\!\!\!\!{\mathfrak A}_1^{(1)}[L_s]
   \!\!&\!\!=\!\!&\!\! \int_s^{\infty}\!\frac{\rho_1(\sigma)}{\sigma}\,d\sigma
            = \frac{1}{\pi}\,\arccos\frac{L_s}{\sqrt{\pi^2+L_s^2}},\!\!\!
 \end{eqnarray}
\end{subequations}
whereas analytic images of the higher powers ($n\geq2, n\in\mathbb{N}$) are:
\begin{eqnarray}
 \label{eq:recurrence}
 {\mathcal A_n^{(1)}[L] \choose \mathfrak A_n^{(1)}[L_s]}
  \!\!&\!\!=\!\!&\!\! \frac{1}{(n-1)!}\left( -\frac{d}{d L}\right)^{n-1}
      {\mathcal A_{1}^{(1)}[L] \choose \mathfrak A_{1}^{(1)}[L_s]}\,.
\end{eqnarray}

\section{Fractional APT in QCD}
\label{sec:FAPT}
At first glance, the APT is a complete theory
providing tools to produce
an analytic answer for any perturbative series in QCD.
But in 2001 Karanikas and Stefanis~\cite{KS01}
suggested the principle of analytization ``as a whole''
in the $Q^2$ plane for hadronic observables,
calculated perturbatively.
More precisely, they proposed the analytization recipe
for terms like
$\int_{0}^{1}\!dx\!\int_{0}^{1}\!dy\,
  \alpha_\text{s}\left(Q^{2}xy\right) f(x)f(y)$,
which can be treated as an effective account
for the logarithmic terms
in the next-to-leading-order
approximation of the perturbative QCD.
This actually generalizes the analytic approach
suggested in~\cite{SSK9900}.
Indeed, in the standard QCD PT one has also:\\
(i) the factorization QCD procedure
    that gives rise to the appearance of logarithmic factors of the type:
     $a^\nu[L]\,L$;\\
(ii) the RG evolution
     that generates evolution factors of the type:
     $B(Q^2)=\left[Z(Q^2)/Z(\mu^2)\right]$ $B(\mu^2)$,
     which reduce in the one-loop approximation to
     $Z(Q^2) \sim a^\nu[L]$ with $\nu=\gamma_0/(2b_0)$
     being a fractional number.\\
All that means
that in order to generalize APT
in the ``analytization as a whole'' direction
one needs to construct analytic images
of new functions:
$\displaystyle a^\nu,~a^\nu\,L^m, \ldots$\,.
This task has been performed in the frames of the so-called FAPT,
suggested in~\cite{BMS-APT,BKS05}.
Now we briefly describe this approach.

In the one-loop approximation
using recursive relation (\ref{eq:recurrence})
we can obtain explicit expressions for
both couplings:
\begin{subequations}
\begin{eqnarray}
 {\mathcal A}_{\nu}^{(1)}[L]
 \!\!&\!\!=\!\!&\!\!
   \frac{1}{L^\nu}
  - \frac{F(e^{-L},1-\nu)}{\Gamma(\nu)}\,;
 ~
 \\
 {\mathfrak A}_{\nu}^{(1)}[L]
 \!\!&\!\!=\!\!&\!\!
   \frac{\text{sin}\left[(\nu -1)\arccos\left(\frac{L}{\sqrt{\pi^2+L^2}}\right)\right]}
         {\pi(\nu -1) \left(\pi^2+L^2\right)^{(\nu-1)/2}}\,.~
\end{eqnarray}
\end{subequations}
Here $F(z,\nu)$ is reduced Lerch transcendental function,
which is an analytic function in $\nu$.
The obtained functions,
${\mathcal A}_{\nu}^{(1)}[L]$
and ${\mathfrak A}_{\nu}^{(1)}[L]$,
have very interesting properties,
which we discussed extensively in our previous papers~\cite{BMS-APT,BKS05,AB08,Ste09}.
Note here,
that in the one-loop approximation
to find analytic images of $a_{(1)}^{\nu}[L]\cdot L^m$
is very easy:
they are just $\mathcal A_{\nu-m}^{(1)}[L]$
and $\mathfrak A_{\nu-m}^{(1)}[L]$.

Constructing FAPT in the higher-($l$)-loop  approximations
is a more complicated task.
Here we represent the original
$l$-loop running coupling
in the following form:
\begin{eqnarray}
 \label{eq:a_(2).1.Lamb.R.phi}
  a_{(l)}\left[L_\sigma-i\pi\right]
   = \frac{\displaystyle e^{i\varphi_{(l)}[L_\sigma]}}{R_{(l)}[L_\sigma]}\,,
\end{eqnarray}
with
$R_{(l)}[L]$
and $\varphi_{(l)}[L]$
being the known functions.
Then the spectral densities of the $\nu$-power of the coupling is
\begin{eqnarray}
 \label{eq:SpDen.nu.(l)}
  \rho_{\nu}^{(l)}[L_\sigma]
  = \frac{1}{\pi}\,
     \frac{\sin[\nu~\varphi_{(l)}[L_\sigma]]}
          {\left(R_{(l)}[L_\sigma]\right)^{\nu}}\,,
\end{eqnarray}
but spectral densities of $a^{\nu}[L]\cdot L^m$
start to be more complicated:
\begin{eqnarray}
 \rho^{(l)}_{{\cal L}_{\nu,m}}[L]
  = \frac{R_{(1)}^m[L]}
         {\pi\,R_{(l)}^\nu[L]}
     \sin\left[\nu\varphi_{(l)}[L]
              - m\varphi_{(1)}[L]
         \right].~
 \label{eq:Sp.Den.L.nu.m.(l)}
\end{eqnarray}

Construction of FAPT with fixed number of quark flavors, $N_f$,
is a two-step procedure:
we start with the perturbative result $\left[a(Q^2)\right]^{\nu}$,
generate the spectral density $\rho_{\nu}(\sigma)$ using Eq.\ (\ref{eq:An.SD}),
and then obtain analytic couplings
${\mathcal A}_{\nu}[L]$ and ${\mathfrak A}_{\nu}[L]$ via Eqs.\ (\ref{eq:A.U}).
Here $N_f$ is fixed and factorized out.
We can proceed in the same manner for $N_f$-dependent quantities:
$\left[\alpha_s^{}(Q^2;N_f)\right]^{\nu}$
$\Rightarrow$
$\bar{\rho}_{\nu}(\sigma;N_f)=\bar{\rho}_{\nu}[L_\sigma;N_f]
 \equiv\rho_{\nu}(\sigma)/\beta_f^{\nu}$
$\Rightarrow$
$\bar{\mathcal A}_{\nu}^{}[L;N_f]$ and $\bar{\mathfrak A}_{\nu}^{}[L;N_f]$ ---
here $N_f$ is fixed, but not factorized out.

Global version of FAPT~\cite{AB08},
which takes into account heavy-quark thresholds,
is constructed along the same lines
but starting from global perturbative coupling
$\left[\alpha_s^{\,\text{\tiny glob}}(Q^2)\right]^{\nu}$,
being a continuous function of $Q^2$
due to choosing different values of QCD scales $\Lambda_f$,
corresponding to different values of $N_f$.
We illustrate here the case of only one heavy-quark threshold
at $s=m_4^2$,
corresponding to the transition $N_f=3\to N_f=4$.
Then we obtain the discontinuous spectral density
\begin{eqnarray}
 \rho_n^\text{\tiny glob}(\sigma)
  \!\!&\!\!=\!\!&\!\! \theta\left(L_\sigma<L_{4}\right)\,
       \bar{\rho}_n\left[L_\sigma;3\right]
  \nonumber\\
  \!\!&\!\!+\!\!&\!\! \theta\left(L_{4}\leq L_\sigma\right)\,
       \bar{\rho}_n\left[L_\sigma+\lambda_4;4\right]\,,~
 \label{eq:global_PT_Rho_4}
\end{eqnarray}
with $L_{\sigma}\equiv\ln\left(\sigma/\Lambda_3^2\right)$,
$L_{f}\equiv\ln\left(m_f^2/\Lambda_3^2\right)$
and
$\lambda_f\equiv\ln\left(\Lambda_3^2/\Lambda_f^2\right)$ for $f=4$,
which is expressed in terms of fixed-flavor spectral densities
with 3 and 4 flavors,
$\bar{\rho}_n[L;3]$ and $\bar{\rho}_n[L+\lambda_4;4]$.
However it generates the continuous Minkowskian coupling
\begin{subequations}
\begin{eqnarray}
 {\mathfrak A}_{\nu}^{\text{\tiny glob}}[L]
  \!\!&\!\!=\!\!&\!\!
    \theta\left(L\!<\!L_4\right)
     \Bigl(\bar{{\mathfrak A}}_{\nu}^{}[L;3]
          + \Delta_{43}\bar{{\mathfrak A}}_{\nu}^{}
     \Bigr)
  \nonumber\\
  \!\!&\!\!+\!\!&\!\!
    \theta\left(L_4\!\leq\!L\right)\,
     \bar{{\mathfrak A}}_{\nu}^{}[L+\lambda_4;4]\,.
 \label{eq:An.U_nu.Glo.Expl}
\end{eqnarray}
with $\Delta_{43}\bar{{\mathfrak A}}_{\nu}^{}=
            \bar{{\mathfrak A}}_{\nu}^{}[L_4+\lambda_4;4]
          - \bar{{\mathfrak A}}_{\nu}^{}[L_4;3]
$
and the analytic Euclidean coupling ${\cal A}_{\nu}^{\text{\tiny glob}}[L]$
\begin{eqnarray}
 \!\!\!\!\!\!\!{\cal A}_{\nu}^{\text{\tiny glob}}[L]
  \!\!&\!\!=\!\!&\!\! \bar{{\cal A}}_{\nu}^{}[L+\lambda_4;4]
 \nonumber\\
  \!\!\!\!\!\!\!
  \!\!&\!\!+\!\!&\!\! \int\limits_{-\infty}^{L_4}\!
       \frac{\bar{\rho}_{\nu}^{}[L_\sigma;3]
            -\bar{\rho}_{\nu}^{}[L_\sigma+\lambda_{4};4]}
            {1+e^{L-L_\sigma}}\,
         dL_\sigma
  \label{eq:Delta_f.A_nu}
\end{eqnarray}
\end{subequations}
(for more detail see in~\cite{AB08}).

\section{Resummation in (F)APT}
\label{sec:Resum.FAPT}
  Before starting with the definite-loop approximation
we introduce here,
following~\cite{MS04},
the generating function $P(t)$
for perturbative coefficients $d_k$
which allows us then to resum any non-power series
($\nu=0$ corresponds to the APT case) of the type
\begin{eqnarray}
 \mathcal S_\nu[L;\mathcal F]
  = d_0\,\mathcal F_{\nu}[L]
   + d_1\,\sum_{n\geq1}
      \tilde{d}_{n}\,
      \mathcal F_{n+\nu}[L]\,,
\label{eq:series.S.FAPT}
\end{eqnarray}
where
${\mathcal F}[L]$ denotes one of the analytic quantities
$\mathcal A^{(l)}[L]$,
$\mathfrak A^{(l)}[L]$,
or $\rho^{(l)}[L]$,
and
$\tilde{d}_n\equiv d_n/d_1$.
We suppose
that
\begin{eqnarray}
  \tilde{d}_n
=
  \int_{0}^\infty\!\!P(t)\,t^{n-1}dt
  ~~~\text{with}~~~
  \int_{0}^\infty\!\!P(t)\,d t = 1\, .
\label{eq:generator}
\end{eqnarray}
To shorten our formulas
we use the abbreviated notation
\begin{eqnarray}
 \left\langle\!\left\langle{F[L,t]}
 \right\rangle\!\right\rangle_{P(t)}
 \equiv
  \int^{\infty}_{0}F[L,t]\,P(t)~dt\,.
\label{eq:convolution}
\end{eqnarray}
Then $\tilde{d}_n=\left\langle\!\left\langle{t^{n-1}}
      \right\rangle\!\right\rangle_{P(t)}$
and we need to resum the series
\begin{subequations}
\begin{eqnarray}
 \mathcal W_\nu[L;t;\mathcal F]
  = \sum_{n\geq1}
      t^{n-1}\,
      \mathcal F_{n+\nu}[L]\,,
\label{eq:series.W.FAPT}
\end{eqnarray}
related to the original one in a simple way
\begin{eqnarray}
 \!\!\!\!\!\mathcal S_\nu[L;\mathcal F]
 = d_0\,\mathcal F_{\nu}[L]
   + d_1
   \left\langle\!\left\langle{\mathcal W_\nu[L;t;\mathcal F]}
      \right\rangle\!\right\rangle_{P(t)}\,.~
 \label{eq:series.S.FAPT.P(t)}
\end{eqnarray}
\end{subequations}

\subsection{One-loop FAPT}
\label{sec:Resum.FAPT.1L}
In the one-loop approximation
we have the following recurrence relation
($\dot{\mathcal F}[L]\equiv d\mathcal F[L]/dL$):
\begin{eqnarray}
  -\frac{1}{n+\nu}\,
    \dot{\mathcal F}_{n+\nu}[L]
  = \mathcal F_{n+1+\nu}[L]\,.
\label{eq:rec.rel.FAPT.1L}
\end{eqnarray}
This property of couplings and spectral densities
allows us to resum the series (\ref{eq:series.S.FAPT.P(t)}):
\begin{subequations}
\label{eq:FAPT.Sum.Appro.1L}
\begin{eqnarray}
 \!\!\!\!\!\!\! \mathcal S_\nu[L,\mathcal F]
  = d_0\,\mathcal F_\nu[L]
  + \hat{d}_1 \left\langle\!\left\langle{{\mathcal F}_{1+\nu}[L-t]}
                           \right\rangle\!
              \right\rangle_{P_\nu(t)}\,,
 \label{eq:FAPT.Sum.1L}
\end{eqnarray}
where now the generating function $P_\nu$ depends on $\nu$,
\begin{eqnarray}
  P_{\nu}(t)
=
  \int_0^{1}\!P\left(\frac{t}{1-x}\right)
  \Phi_{\nu}(x) \frac{dx}{1-x}\,.
\label{eq:P.nu}
\end{eqnarray}
Here
$\Phi_{\nu}(x) = \nu x^{\nu-1}$, so that
$\lim_{\nu\to 0}\Phi_{\nu}\rightarrow \delta(x)$, and therefore
$\lim_{\nu\to 0}P_{\nu}(t)= P(t)$.
\end{subequations}

\subsection{Two-loop FAPT}
\label{sec:Resum.FAPT.2L}
In the two-loop approximation
we have more complicated recurrence relation
\begin{eqnarray}
  -\frac{1}{n+\nu}\,
    \dot{\mathcal F}_{n+\nu}[L]
  = \mathcal F_{n+1+\nu}[L]
  + c_1\,\mathcal F_{n+2+\nu}[L]
\label{eq:rec.rel.FAPT.2L}
\end{eqnarray}
with $c_1$ being the corresponding coefficient in Eq.\,(\ref{eq:RG.a}).
In order to resum the series $\mathcal W_\nu[L;t;\mathcal F]$
we need to introduce the ``two-loop evolution'' time
\begin{eqnarray}
 \label{eq:2L.Evo.Time}\!\!\!\!\!\!
  \tau_2(t)
   = t
   - c_1\,\ln\left[1+ \frac{t}{c_1}\right]\,;
 \
 \dot{\tau}_2(t)
  = \frac{1}{1+c_1/t}\,.~
\end{eqnarray}
We obtained in~\cite{BMS10} the following resummation
recipe
\begin{eqnarray}
 \label{eq:FAPT.Sum.2L}
  \!\!\!\!\mathcal W_\nu[L,t;\mathcal F]
  = \mathcal F_{\nu+1}[L]
  + \Delta(\nu)\,
     c_1\,\dot{\tau}_2(t)\,
      \mathcal F_{2}[L_{t,0}]
\nonumber\\ \!\!\!\!\!\!\!\!\!\!
  - \dot{\tau}_2(t)
     \int\limits_0^1\!
      z^{\nu}
       \Big[t\,\dot{\mathcal{F}}_{\nu+1}[L_{t,z}]
           - \frac{c_1\,\nu}{z}\,\mathcal{F}_{\nu+2}[L_{t,z}]
       \Big]\,dz~
\end{eqnarray}
with
$L_{t,z}=L+\tau_{2}(t\,z)-\tau_{2}(t)$,
$L_{t,0}=L-\tau_{2}(t)$,
and
$\Delta(\nu)$ being a Kronecker delta symbol.
Interesting to note here
that
it is possible to obtain an analogous,
but more complicated recipe
for the case when
$\mathcal F$ is the analytic image
of the two-loop evolution factor
$a^\nu(1+c_1 a)^{\nu_1}$,
see in~\cite{BMS10} for more detail.

\begin{table*}[!t]
 \centering
 \caption{Coefficients $d_n$ for the Adler-function series
with $N_f=4$.
The numbers in the square brackets denote the lower and the upper
limits of the INNA estimates.
\label{Tab:d_n.Adler}\vspace*{+1.3mm}}
{\small
\begin{tabular}{|c|c|ccccc|}\hline \hline
                  & PT coefficients
                       &~$d_1\vphantom{^{|}_{|}}$~
                             &~$d_2$~
                                    &~$d_3$~
                                           &~$d_4$~
                                                  &~$d_5$~
\\ \hline \hline
$1^{\phantom{+}}$ &~pQCD results with $N_f=4$ \cite{BCK08,BCK10}~
                       & $1\vphantom{^{|}_{|}}$
                             &~1.52~&~2.59~&~27.4~&~---~
\\ \hline \hline
$2^{+}$           &~Model (\ref{eq:Vector.Model}) with $c=3.544,~\delta=1.3252$~
                       & $1\vphantom{^{|}_{|}}$
                             &~1.53~&~2.80~&~30.9~&~2088~
\\ \hline
$2^{\phantom{+}}$ &~Model (\ref{eq:Vector.Model}) with $c=3.553,~\delta=1.3245$~
                       & $1\vphantom{^{|}_{|}}$
                             &~1.52~&~2.60~&~27.3~&~2025~
\\ \hline
$2^{-}$           &~Model (\ref{eq:Vector.Model}) with $c=3.568,~\delta=1.3238$~
                       & $1\vphantom{^{|}_{|}}$
                             &~1.52~&~2.39~&~23.5~&~1969~
\\ \hline
$3^{\phantom{+}}$ &~``INNA'' prediction of~~\cite{BMS10}~
                       & $1\vphantom{^{|}_{|}}$
                             &~1.44~&~$[3.5,9.6]$~
                                            &~$[20.4,48.1]$~
                                                   &~$[674,2786]$~
\\ \hline \hline
\end{tabular}}
\end{table*}

\subsection{Three-loop FAPT}
\label{sec:Resum.FAPT.3L}
 We describe here the recently obtained results
on resummation in the three-loop FAPT.
In this case the recurrence relation has three terms in the r.h.s.
\begin{eqnarray}
 \!\!\!\!\!\!\!-\frac{1}{n+\nu}\,
  \dot{\mathcal F}_{n+\nu}[L]
   \!\!&\!=\!&\!\! \mathcal F_{n+1+\nu}[L]
\nonumber\\
  \!\!\!\!\!\!\!
   \!\!&\!+\!&\!\! c_1\,\mathcal F_{n+2+\nu}[L]
          + c_2\,\mathcal F_{n+3+\nu}[L]
\label{eq:rec.rel.FAPT.3L}
\end{eqnarray}
with $c_2$ being the corresponding coefficient in Eq.\,(\ref{eq:RG.a}).
We introduce the ``three-loop evolution'' time
by
\begin{eqnarray}
 \label{eq:time.3L}
 \tau_3(t)
  \!\!&\!=\!&\!\! t
  + \frac{c_1^2-2 c_2}{\Delta}
     \arctan\left[\frac{t\,\Delta}{2\,c_2+c_1\,t}\right]
 \nonumber\\
  \!\!&\!-\!&\!\!
    \frac{c_1}{2}\,
     \ln\left[1+\frac{c_1+t}{c_2}\,t\right]\,;
 \\ \nonumber
 \frac{d\tau_3(t)}{d\,t}
 \!\!&\!\equiv\!&\!\!
  \dot{\tau}_3(t) =
    \frac{1}{1+c_1/t+c_2/t^2}\,.~
\end{eqnarray}
Then our resummation recipe is
\begin{eqnarray}
 \label{eq:FAPT.Sum.3L}
  \!\!\!\!\!\!\!\!\!\!\!\!\!
  \mathcal W_\nu[L,t;\mathcal F]
  = \mathcal F_{\nu+1}[L]
   +  \Delta(\nu)\,c_2\,\dot{\tau}_3(t)\,
      \mathcal F_{3}[L_{t,0}]\!\!\!\!\!
\nonumber\\ \!\!\!\!\!\!\!\!\!\!\!\!\!
   + t\,\mathcal F_{\nu+2}[L]
   - \dot{\tau}_3(t)\!\!
      \int\limits_0^1\!
       z^{\nu}
       \bigg[t\,\dot{\mathcal F}_{\nu+1}[L_{t,z}]
        - \frac{c_2\,\nu}{z}\,\mathcal F_{\nu+3}[L_{t,z}]\!\!\!\!\!
\nonumber\\ \!\!\!\!\!\!\!\!\!\!\!\!
        + z\,t^2\,\dot{\mathcal F}_{\nu+2}[L_{t,z}]
        + (\nu+1)\,t\,{\mathcal F}_{\nu+2}[L_{t,z}]
       \bigg]\,dz\!\!\!\!\!
\end{eqnarray}
with
$L_{t,z}=L+\tau_{3}(t\,z)-\tau_{3}(t)$ and
$L_{t,0}=L-\tau_{3}(t)$.

\section{Resummation for Adler function}
\label{sec:Resum.Adler}
Here we consider the power series
of the vector correlator Adler function
(labeled by the symbol V)~\cite{BCK08,BCK10}
\begin{eqnarray}
  D_\text{V}[L]
  = 1
   + \sum_{n\geq1}d_{n}\,
      \left(\frac{\alpha_s[L]}{\pi}\right)^{n}\,.
 \label{eq:D_V}
\end{eqnarray}
Due to $d_1=1$
coefficients $\tilde{d}_{n}$
coincide with $d_n$.
We suggested~\cite{BMS10}
the model for the generating function
of the perturbative coefficients $d_n$
(see 1st row in Table \ref{Tab:d_n.Adler})
\begin{subequations}
\label{eq:Vector.Model}
 \begin{eqnarray}
 \label{eq:Vector.P(t).Model}
  P_\text{V}(t)
  = \frac{\delta\,e^{-t/c\delta} - (t/c)\,e^{-t/c}}
         {c\left(\delta ^2-1\right)}\,,~
\end{eqnarray}
which provides the following Lipatov-like coefficients
\begin{eqnarray}
 \label{eq:Vector.d_n.Model}
  d_n^\text{V}
  = c^{n-1}\,
    \frac{\delta^{n+1}-n}
    {\delta^2-1}\,\Gamma(n)\,.
\end{eqnarray}
\end{subequations}
Our prediction $d_4^\text{V}=27.1$,
obtained with this generating function by fitting
the two known coefficients $d_2$ and $d_3$
and using the model (\ref{eq:Vector.Model}),
is in a good agreement
with the value 27.4,
calculated in Ref.\ \cite{BCK08,BCK10}.
Note that fitting procedure,
taking into account the fourth-order coefficient $d_4$,
produces the readjustment of the model parameters
in (\ref{eq:Vector.Model}) to the new values
$\left\{c=3.5548,~\delta=1.32448\right\}$
$\to$
$\left\{c=3.5526,~\delta=1.32453\right\}$.
The corresponding values of coefficients $d_n^\text{V}$
are shown in the third row labelled by $2$
in Table~\ref{Tab:d_n.Adler}.

In order to understand how important
are the exact values
of the higher-order coefficients $d_n$,
we employed our model (\ref{eq:Vector.Model})
with two different sets of parameters
$c$ and $\delta$,
shown in rows labelled by $2^{+}$ and $2^{-}$
in Table~\ref{Tab:d_n.Adler}.
One set, $2^+$, roughly speaking, enhances the exact values
of the coefficients $d_3$ and $d_4$ by approximately +8\% and +13\%,
correspondingly,
while the other one, $2^-$, --- reduces them in the same proportion.
All coefficients of these models
are inside the range of uncertainties
determined in~\cite{BMS10} using the Improved Naive Non-Abelinization (INNA).
Moreover,
the difference between the analytic sums of the two models in the
region corresponding to $N_f=4$ is indeed very small, reaching just
a mere $\pm0.05\%$.
This gives an evident support for our model evaluation.

\begin{figure}[h!]
\centerline{\includegraphics[width=0.45\textwidth]{
 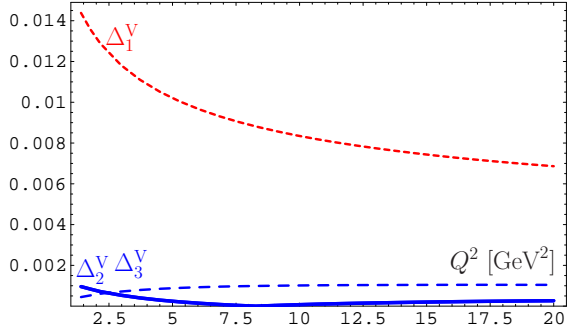} \vspace*{-1mm}}
   \caption{The relative errors $\Delta^\text{V}_N(Q^2)$ evaluated for
    different values of $N$:
    $N=1$ (short-dashed red line),
    $N=2$ (solid blue line), and
    $N=3$ (dashed blue line) of the truncated APT given by
    Eq.\,(\ref{eq:Trun.Err.V}), in comparison with the exact
    result of the one- and two-loop resummation procedure
    represented by Eqs.\,(\ref{eq:FAPT.Sum.Appro.1L})
    and (\ref{eq:FAPT.Sum.2L}).
   \label{fig:Adler}}\vspace*{-1mm}
\end{figure}

Now we are ready to estimate the relative errors,
$\Delta^\text{V}_N[L]$,
of the APT series\footnote{
Note that power series (\ref{eq:D_V})
has $\nu=0$ --- for this reason we use here
the APT approach.} truncation at the $N$th term:
\begin{subequations}
\begin{eqnarray}
 \label{eq:Trun.Adler}
  \mathcal D^\text{V}_{N}[L]
   \!\!&\!=\!&\!\!
    1 + \sum_{n\geq1}^{N}
        \frac{d_{n}}{\pi^{n}}\,\bar{\mathcal A}_n[L]\,;
\\
 \label{eq:Trun.Err.V}
  \Delta^\text{V}_N[L]
   \!\!&\!=\!&\!\!
    \frac{\mathcal D^\text{V}_{\infty}[L]-\mathcal D^\text{V}_{N}[L]}
         {\mathcal D^\text{V}_{\infty}[L]}\,.
\end{eqnarray}
\end{subequations}
Here $\mathcal D^\text{V}_{\infty}[L]$
is the resummed  APT result in the corresponding
$l$-loop approximation,
see Eqs.\,(\ref{eq:FAPT.Sum.1L}), (\ref{eq:FAPT.Sum.2L}), and (\ref{eq:FAPT.Sum.3L})
with substitution $\nu\to0$.
In Fig.\,\ref{fig:Adler}
we show these relative errors for $N=1,\, 2,\, 3$,
for the one- and two-loop cases
(calculations for the three-loop case is not yet finished).
The main result is in some sense surprising:
The best order of truncation of the FAPT series
in the region $Q^2=2-20$~GeV$^2$
is reached by employing the N$^2$LO approximation,
i.e., by keeping just the $d_2$-term.

We may also compare
the numerical values for the resummed quantities,
obtained in different loop approximations.
We take for this comparison
the following $l$-loop QCD scale parameters at $N_f=3$ flavors:
$\Lambda_{3}^{(l=1)}=201$~MeV,
$\Lambda_{3}^{(l=2)}=379$~MeV,
and
$\Lambda_{3}^{(l=3)}=385$~MeV,
which have been determined from the condition
that the APT prediction
for the ratio $R_{e^+e^-}(s=m_Z^2)$
should coincide with the ``experimental'' value $1.03904$,
determined in~\cite{BCK10}.
We obtain the following values of the resummed Adler functions,
shown in the table form for two values of $Q^2$,
namely 3 and 2\,GeV$^2$:
\vspace*{+1.5mm}

\centerline{\begin{tabular}{c|ccc}\hline
  Loop order $l$ & $l=1$   & $l=2$   & $l=3\vphantom{^{\int}}$   \\ \hline \hline
  $\mathcal D^\text{V, $(l)$}_{\infty}(Q^2=3\,\text{GeV}^2)\vphantom{^{\int}_|}$
                 & 1.1129 & 1.1131 & 1.1164 \\ \hline
  $\mathcal D^\text{V, $(l)$}_{\infty}(Q^2=2\,\text{GeV}^2)\vphantom{^{\int}_|}$
                 & 1.1221 & 1.1223 & 1.1257 \\ \hline
\end{tabular}}\vspace*{+1.5mm}

\noindent Strictly speaking,
our model for coefficients is valid only
for $Q^2\gtrsim m_c^2=2.44$\,GeV$^2$,
so that the first value $Q^2=3$\,GeV$^2$ was selected
to show the results
in the legitimate $N_f=4$ domain.
The second value, $Q^2=2$\,GeV$^2$, was selected
for the comparison
with the recent estimate in~\cite{CvKo11},
where the value
$\mathcal D^\text{V}_{\infty}(Q_2^2)=1.1217$
has been obtained in the two-loop approximation
using the so-called generalized Pade summation method.

\begin{table*}[!t]
 \centering
 \caption{Coefficients $d_n$ for the Higgs-boson-decay width series with $N_f=5$.
  \label{Tab:d_n.Higgs}\vspace*{+1.3mm}}
{\small
\begin{tabular}{|c|c|ccccc|}\hline \hline
                  & PT coefficients
                       &~$\tilde{d}_1\vphantom{^{\big|}_{|}}$~
                             &~$\tilde{d}_2$~
                                    &~$\tilde{d}_3$~
                                           &~$\tilde{d}_4$~
                                                 &~$\tilde{d}_5$~
\\ \hline \hline
$1^{\phantom{+}}$ &~pQCD results with $N_f=4$ \cite{BCK05}~
                       & $1\vphantom{^{|}_{|}}$
                             &~7.42~&~62.3~&~620~&~---~
\\ \hline \hline
$2^{+}$           &~Model (\ref{eq:Higgs.Model}) with $c=2.43,~\delta=-0.52$~
                       & $1\vphantom{^{|}_{|}}$
                             &~7.85~&~68.5~&~752~& 10120~
\\ \hline
$2^{\phantom{+}}$ &~Model (\ref{eq:Higgs.Model}) with $c=2.62,~\delta=-0.50$~
                       & $1\vphantom{^{|}_{|}}$
                             &~7.50~&~61.1~&~625~&~7826~
\\ \hline
$2^{-}$           &~Model (\ref{eq:Higgs.Model}) with $c=2.25,\delta=-0.51$~
                       & $1\vphantom{^{|}_{|}}$
                             &~6.89~&~52.0~&~492~&~5707~
\\ \hline
$3^{\phantom{+}}$ &~``PMS'' predictions of~\cite{KaSt95,ChKS97,BCK05}~
                       &~$-\vphantom{^{|}_{|}}$~
                             &~$-$~ &~64.8~&~547~&~7782~
\\ \hline \hline
\end{tabular}}
\end{table*}

\section{Resummation for $H^0\to\bar{b}b$ Decay Width}
\label{sec:Resum.Higgs}
Here we analyze the Higgs boson decay to a $\bar{b}b$ pair.
For its width we have
\begin{eqnarray}
 \label{eq:Higgs.decay.rate}
 \Gamma(\text{H} \to b\bar{b})
  = \frac{G_F}{4\sqrt{2}\pi}\,
     M_{H}\,
      \widetilde{R}_\text{\tiny S}(M_{H}^2)
\end{eqnarray}
with
$\widetilde{R}_\text{\tiny S}(M_{H}^2)
  \equiv m^2_{b}(M_{H}^2)\,R_\text{\tiny S}(M_{H}^2)$
and
$R_\text{\tiny S}(s)$
is the $R$-ratio for the scalar correlator,
see for details in~\cite{BMS-APT,BCK05}.
In the one-loop FAPT this generates the following
non-power expansion:
\begin{eqnarray}
 \!\!\!\!\!\!\!
 \widetilde{\mathcal R}_\text{\tiny S}[L]
   =  3\,\hat{m}_{(1)}^2\,
      \Bigg\{\mathfrak A_{\nu_{0}}^{\text{\tiny glob}}[L]
          + d_1^\text{\,\tiny S}\,\sum_{n\geq1}
             \frac{\tilde{d}_{n}^\text{\,\tiny S}}{\pi^{n}}\,
              \mathfrak A_{n+\nu_{0}}^{\text{\tiny glob}}[L]
      \Bigg\}\,,
 \label{eq:R_S-MFAPT}
\end{eqnarray}
where $\hat{m}_{(1)}=8.21-8.53$~GeV is the RG-invariant
of the one-loop $m_{b}(\mu^2)$ evolution
$m_{b}^2(Q^2) = \hat{m}_{(1)}^2\,\alpha_{s}^{\nu_{0}}(Q^2)$
with $\nu_{0}=2\gamma_0/b_0(5)=1.04$ and
$\gamma_0$ is the quark-mass anomalous dimension
(for a discussion --- see in~\cite{BMS10}).\footnote{
Different values of $\hat{m}_{(1)}$ is related with two different extractions
of the RG-effective $b$-quark mass,
$\overline{m}_b(\overline{m}_b^2)$,
which we have taken from two independent analyses.
One value originates from Ref.\ \cite{PeSt02}
($\overline{m}_b(\overline{m}_b^2)=4.35\pm0.07$~GeV),
while the other was derived in Ref.\ \cite{KuSt01}
yielding $\overline{m}_b(\overline{m}_b^2)=4.19\pm0.05$~GeV.}

We take for the generating function $P(t)$
the model of~\cite{BM08}
with $\left\{c=2.4,~\beta=-0.52\right\}$
\begin{subequations}
\label{eq:Higgs.Model}
\begin{eqnarray}
 \label{eq:Higgs.Model.P}
  P_\text{\tiny S}(t)
  = \frac{(t/c)+\beta}{c\,(1+\beta)}\,e^{-{t/c}}\,.
\end{eqnarray}
It provides the following Lipatov-like coefficients
\begin{eqnarray}
 \label{eq:Higgs.Model.d_n}
  \tilde{d}_{n}^\text{\,\tiny S}
  = c^{n-1}\frac{\Gamma (n+1)+\beta\,\Gamma (n)}{1+\beta}
\end{eqnarray}
\end{subequations}
which are in a very good agreement with
$\tilde{d}_{n}^\text{\,\tiny S}, n=2,\,3,\,4$,
calculated in the QCD PT~\cite{BCK05},
the corresponding values of coefficients $\tilde{d}_n^\text{\,\tiny S}$
are shown in the third row labelled by $2$
in Table~\ref{Tab:d_n.Higgs}.
In order to estimate the importance
of the higher-order coefficients $d_n$ exact values,
we proceed along the same lines
as in Sect.~\ref{sec:Resum.Adler}:
We employ our model (\ref{eq:Higgs.Model})
with two different sets of parameters
$c$ and $\beta$,
shown in rows labelled by $2^{+}$ and $2^{-}$
in Table~\ref{Tab:d_n.Higgs}.
One set, $2^+$, enhances the exact values
of the coefficients $d_3$ and $d_4$ by approximately +13\% and +20\%,
correspondingly,
while the other one, $2^-$, --- reduces them in the same proportion.
Obtained in this way difference between the analytic sums of the two models
in the region corresponding to $M_H=80-170$~GeV
is indeed very small,
not more than $0.5\%$.
Note here also that the model prediction for $d_5$
is very close to the prediction
obtained using the Principle of Minimal Sensitivity (PMS)~\cite{KaSt95},
shown in the row with label $3$
in Table~\ref{Tab:d_n.Higgs}.

After verification of our model quality
we apply the FAPT resummation technique to estimate
how good is FAPT
in approximating the whole sum $\widetilde{\mathcal R}_\text{\tiny S}[L]$
in the range $L\in[12.4,13.5]$
which corresponds to the range
$M_H\in[100,170]$~GeV$^2$
with $\Lambda_{N_f=3}^{\text{QCD}}=201$~MeV.
Here we need to use our resummation recipes
(\ref{eq:FAPT.Sum.1L}) and (\ref{eq:FAPT.Sum.2L})
with substitution $\nu\to\nu_0=1.04$.
Note that in the two-loop approximation
the one-loop evolution factor $a^\nu$ transforms
into a more complicated expression, $a^{\nu}(1+c_1\,a)^{\nu_1}$.
This produces additional numerical complications,
but qualitatively results are the same.
For this reason we show explicitly only one-loop formulas.

We analyze the accuracy of the truncated FAPT expressions
\begin{eqnarray}
 \label{eq:FAPT.trunc}\!\!\!\!\!\!\!
 \widetilde{\mathcal R}_\text{\tiny S}[L;N]
   = 3\,\hat{m}_{(1)}^2\!
       \left[{\mathfrak A}_{\nu_{0}}^{\text{\tiny glob}}[L]
           + d_1^\text{\,\tiny S}\,\sum_{n=1}^{N}
              \frac{\tilde{d}_{n}^\text{\,\tiny S}}{\pi^{n}}\,
               {\mathfrak A}_{n+\nu_{0}}^{\text{\tiny glob}}[L]
       \right]\!\!\!
\end{eqnarray}
and compare them with the resummed FAPT result
$\widetilde{\mathcal R}_\text{\tiny S}[L]$
in the corresponding
$l$-loop approximation\footnote{
Here we show the results only for $l=1$ and $l=2$:
Calculations for $l=3$ are not yet finished.}
using relative errors
$\Delta_N[L]=1-\widetilde{\mathcal R}_\text{\tiny S}[L;N]/\widetilde{\mathcal R}_\text{\tiny S}[L]$.
We estimate these errors for $N=2$, $N=3$, and $N=4$
in the analyzed range of $L\in[11,13.8]$
and show that already $\widetilde{\mathcal R}_\text{\tiny S}[L;2]$
gives accuracy of the order of 2.5\%,
whereas $\widetilde{\mathcal R}_\text{\tiny S}[L;3]$
of the order of 1\%.
That means that there is no need to calculate further corrections:
at the level of accuracy of 1\% it is quite enough to take into account
only coefficients up to $d_3$.
This conclusion is stable
with respect to the variation of parameters
of the model $P_\text{\tiny S}(t)$
and is in a complete agreement
with Kataev--Kim conclusion~\cite{KK09}.

\begin{figure}[h!]
\centerline{\includegraphics[width=0.45\textwidth]{
 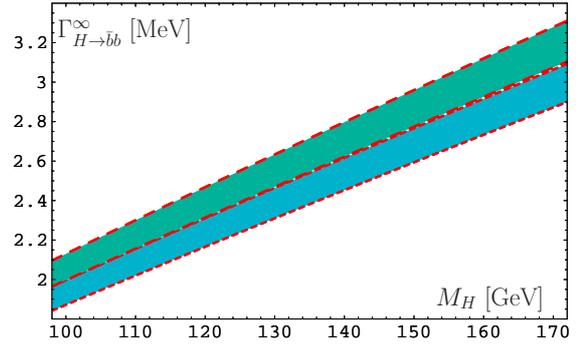} \vspace*{-1mm}}
   \caption{The two-loop width $\Gamma_{H \to b\bar{b}}^{\infty}$
    in the resummed FAPT as a function of the Higgs-boson mass $M_H$.
    The mass is varied in the interval $\hat{m}_{(2)}=8.22\pm0.13$~GeV
    according to the Penin--Steinhauser estimate~\cite{PeSt02}.
    The upper strip shows the corresponding one-loop result.
    \label{fig:Higgs}}\vspace*{-1mm}
\end{figure}
In Fig.\ \ref{fig:Higgs}
we show the results for the decay width
$\Gamma_{H \to b\bar{b}}^{\infty}(M_H)$
in the resummed two-loop FAPT,
in the window of the Higgs mass allowed by
the LEP and Tevatron experiments.
Comparing this outcome with the one-loop result,
shown as the upper strip in Fig.\,\ref{fig:Higgs},
reveals a 5\% reduction of the two-loop estimate.
This reduction consists of two parts:
one part ($\approx+7\%$)
comes from the difference in the mass $\hat{m}$,
while the other ($\approx-2\%$) is due to the difference
in the values of $R_{S}(M_H)$
in both approximations.

\section{Conclusions}
\label{sec:Final}

We conclude with the following.

APT provides natural way to Minkowski region
  for coupling and related quantities with weak loop dependence
  and practical scheme independence.

FAPT provides an effective tool to apply the APT approach
  for the renormalization-group improved perturbative amplitudes.

Both APT and FAPT produce finite resummed answers
(now --- up to the three-loop level)
for perturbative quantities
if we know the generating function $P(t)$
for the PT coefficients.

Using quite simple model generating function $P(t)$
for the Adler function $\mathcal D(Q^2)$
we show that already at the N${^2}$LO
an accuracy is of the order 0.1\%,\\
whereas for the Higgs boson decay $H\to\overline{b}b$
   at the N${^3}$LO is of the order of:
   {1\%} --- due to the truncation error;
   {2\%} --- due to the RG-invariant mass uncertainty.

\newpage

\textbf{Acknowledgements}
We thank our collaborators S.\,V.\,Mikhailov and N.\,G.\,Stefanis
for numerous discussions and useful remarks.
This work was supported by the RFBR grant No. 11-01-00182,
the BRFBR--JINR grant (contract No.~F10D-001)
and the Heisenberg--Landau Program (grant 2011).





\begin{thebibliography}{10}
\expandafter\ifx\csname url\endcsname\relax
  \def\url#1{\texttt{#1}}\fi
\expandafter\ifx\csname urlprefix\endcsname\relax\def\urlprefix{URL }\fi
\expandafter\ifx\csname href\endcsname\relax
  \def\href#1#2{#2} \def\path#1{#1}\fi

\bibitem{KaSh80}
 D.~I. Kazakov, D.~V. Shirkov,
 {Asymptotic series of quantum field theory and their summation},
 Fortsch. Phys. 28 (1980) 465--499.
  \newblock
   \href{http://dx.doi.org/10.1002/prop.19800280803}
              {\path{doi:10.1002/prop.19800280803}}.

\bibitem{Lip76}
 L.~N. Lipatov,
  {Divergence of the Perturbation Theory Series and the Quasiclassical Theory},
   Sov. Phys. JETP 45 (1977) 216--223.

\bibitem{BaSh11}
 A.~P. Bakulev, D.~V. Shirkov,
  {Inevitability and Importance of Non-Perturbative Elements in Quantum Field Theory},
   {Invited talk delivered by the first author at the 6th Mathematical Physics Meeting:
    Summer School and Conference on Modern Mathematical Physics,
    Belgrade, Sept. 15--23, 2010}
    \newblock
    \href{http://arxiv.org/abs/1102.2380}
               {\path{arXiv:1102.2380}} [hep-ph] (2011).

\bibitem{BLS60}
 N.~N. Bogolyubov, A.~A. Logunov, D.~V. Shirkov,
  The method of dispersion relations and perturbation theory,
   Soviet Physics JETP 10 (1960) 574.

\bibitem{Rad82}
 A.~V. Radyushkin,
  {Optimized lambda-parametrization for the QCD running coupling constant
   in space-like and time-like regions},
    JINR Rapid Commun. 78 (1996) 96--99,
    [JINR Preprint, E2-82-159, 26 Febr. 1982].
    \newblock
     \href{http://arxiv.org/abs/hep-ph/9907228}
          {\path{arXiv:hep-ph/9907228}}.

\bibitem{KP82}
 N.~V. Krasnikov, A.~A. Pivovarov,
  {The influence of the analytical continuation effects on the value
   of the QCD scale parameter lambda extracted from the data on charmonium
   and upsilon hadron decays},
    Phys. Lett. B116 (1982) 168--170.

\bibitem{JS95-349}
 H.~F. Jones, I.~L. Solovtsov,
  {QCD running coupling constant in the timelike region},
   Phys. Lett. B349 (1995) 519--524.
   \newblock
    \href{http://arxiv.org/abs/hep-ph/9501344}
               {\path{arXiv:hep-ph/9501344}}.

\bibitem{SS}
 D.~V. Shirkov, I.~L. Solovtsov,
  {Analytic QCD running coupling with finite IR behaviour and universal
   $\bar{\alpha}_s(0)$ value},
    JINR Rapid Commun. 2[76] (1996) 5--10.
    \newblock
     \href{http://arxiv.org/abs/hep-ph/9604363}
          {\path{arXiv:hep-ph/9604363}}.\\
  {Analytic model for the QCD running coupling with universal
   $\bar{\alpha}_s(0)$ value},
    Phys. Rev. Lett. 79 (1997) 1209--1212.
    \newblock
     \href{http://arxiv.org/abs/hep-ph/9704333}
          {\path{arXiv:hep-ph/9704333}}.\\
  {Ten years of the analytic perturbation theory in QCD},
    Theor. Math. Phys. 150 (2007) 132--152.
    \newblock
     \href{http://arxiv.org/abs/hep-ph/0611229}
          {\path{arXiv:hep-ph/0611229}}.

\bibitem{PST}
 R.~S. Pasechnik \textit{et al.}
  {Nucleon spin structure and pQCD frontier on the move},
   Phys. Rev. D81 (2010) 016010.
    \newblock
     \href{http://arxiv.org/abs/0911.3297}
          {\path{arXiv:0911.3297}},
     \href{http://dx.doi.org/10.1103/PhysRevD.81.016010}
          {\path{doi:10.1103/PhysRevD.81.016010}}.

\bibitem{KS01}
 A.~I. Karanikas, N.~G. Stefanis,
  {Analyticity and power corrections in hard-scattering hadronic functions},
   Phys. Lett. B504 (2001) 225--234;
   Erratum --- \textit{ibid.} B636 (2006) 330.
   \newblock
    \href{http://arxiv.org/abs/hep-ph/0101031}
         {\path{arXiv:hep-ph/0101031}}.

\bibitem{SSK9900}
 N.~G. Stefanis, W.~Schroers, H.-C. Kim,
  {Pion form factors with improved infrared factorization},
   Phys. Lett. B449 (1999) 299.
   \newblock
    \href{http://arxiv.org/abs/hep-ph/9807298}
         {\path{arXiv:hep-ph/9807298}}.\\
  {Analytic coupling and Sudakov effects in exclusive processes:
   Pion and $\gamma^*\gamma\to\pi^0$ form factors},
   Eur. Phys. J. C18 (2000) 137--156.
   \newblock
    \href{http://arxiv.org/abs/hep-ph/0005218}
         {\path{arXiv:hep-ph/0005218}}.

\bibitem{BMS-APT}
 A.~P. Bakulev, S.~V. Mikhailov, N.~G. Stefanis,
  {QCD analytic perturbation theory: From integer powers to any power
   of the running coupling},
   Phys. Rev. D72 (2005) 074014, 119908(E).
   \newblock
    \href{http://arxiv.org/abs/hep-ph/0506311}
         {\path{arXiv:hep-ph/0506311}}.\\
  {Fractional analytic perturbation theory in Minkowski space
   and application to Higgs boson decay into a $b\bar{b}$ pair},
    Phys. Rev. D75 (2007) 056005; D77 (2008) 079901(E).
   \newblock
    \href{http://arxiv.org/abs/hep-ph/0607040}
         {\path{arXiv:hep-ph/0607040}}.

\bibitem{BKS05}
 A.~P. Bakulev, A.~I. Karanikas, N.~G. Stefanis,
  {Analyticity properties of three-point functions in QCD beyond leading order},
   Phys. Rev. D72 (2005) 074015.
   \newblock
    \href{http://arxiv.org/abs/hep-ph/0504275}
         {\path{arXiv:hep-ph/0504275}}.

\bibitem{AB08}
 A.~P. Bakulev,
  {Global Fractional Analytic Perturbation Theory in QCD with Selected Applications},
   Phys. Part. Nucl. 40 (2009) 715--756.
   \newblock
    \href{http://arxiv.org/abs/arXiv:0805.0829 [hep-ph]}
         {\path{arXiv:0805.0829 [hep-ph]}}.

\bibitem{Ste09}
 N.~G. Stefanis,
  {Taming Landau singularities in QCD perturbation theory: The analytic approach},
   \newblock
    \href{http://arxiv.org/abs/0902.4805}
         {\path{arXiv:0902.4805}}  [hep-ph] (2009).

\bibitem{MS04}
 S.~V. Mikhailov,
  {Generalization of BLM procedure and its scales in any order of pQCD:
   A practical approach},
   JHEP 06 (2007) 009.
   \newblock
    \href{http://arxiv.org/abs/hep-ph/0411397}
         {\path{arXiv:hep-ph/0411397}}.

\bibitem{BMS10}
 A.~P. Bakulev, S.~V. Mikhailov, N.~G. Stefanis,
  {Higher-order QCD perturbation theory in different schemes:
   From FOPT to CIPT to FAPT},
   JHEP 1006 (2010) 085.
   \newblock
    \href{http://arxiv.org/abs/1004.4125}
         {\path{arXiv:1004.4125}},
    \href{http://dx.doi.org/10.1007/JHEP06(2010)085}
         {\path{doi:10.1007/JHEP06(2010)085}}.

\bibitem{BCK08}
 P.~A. Baikov, K.~G. Chetyrkin, J.~H. K{\"u}hn,
  {Order $\alpha^4_s$ QCD Corrections to $Z$ and $\tau$ Decays},
   Phys. Rev. Lett. 101 (2008) 012002.
    \newblock
     \href{http://arxiv.org/abs/0801.1821}
          {\path{arXiv:0801.1821}},
     \href{http://dx.doi.org/10.1103/PhysRevLett.101.012002}
          {\path{doi:10.1103/PhysRevLett.101.012002}}.

\bibitem{BCK10}
 P.~A. Baikov, K.~G. Chetyrkin, J.~H. Kuhn,
  {Higgs Function, Bjorken Sum Rule, and the Crewther Relation
   to Order $\alpha_s^4$ in a General Gauge Theory},
   Phys. Rev. Lett. 104 (2010) 132004.
   \newblock
    \href{http://arxiv.org/abs/1001.3606}
         {\path{arXiv:1001.3606}},
    \href{http://dx.doi.org/10.1103/PhysRevLett.104.132004}
         {\path{doi:10.1103/PhysRevLett.104.132004}}.

\bibitem{CvKo11}
 G.~Cvetic, R.~Kogerler,
  {Applying generalized Pad\'e approximants in analytic QCD models},
   \newblock
    \href{http://arxiv.org/abs/1107.2902}
         {\path{arXiv:1107.2902}} [hep-ph] (2011).

\bibitem{BCK05}
 P.~A. Baikov, K.~G. Chetyrkin, J.~H. K{\"u}hn,
  {Scalar correlator at $O(\alpha_s^4)$, Higgs decay into $b$-quarks
   and bounds on the light quark masses},
    Phys. Rev. Lett. 96 (2006) 012003.
    \newblock
     \href{http://arxiv.org/abs/hep-ph/0511063}
          {\path{arXiv:hep-ph/0511063}}.

\bibitem{KaSt95}
 A.~L. Kataev, V.~V. Starshenko,
  {Estimates of the higher order QCD corrections to $R(s)$, $R_\tau$ and
   deep inelastic scattering sum rules},
    Mod. Phys. Lett. A10 (1995) 235--250.
    \newblock
     \href{http://arxiv.org/abs/hep-ph/9502348}
          {\path{arXiv:hep-ph/9502348}},
     \href{http://dx.doi.org/10.1142/S0217732395000272}
          {\path{doi:10.1142/S0217732395000272}}.

\bibitem{ChKS97}
 K.~G. Chetyrkin, B.~A. Kniehl, A.~Sirlin,
  {Estimations of order $\alpha_s^3$ and $\alpha_s^4$ corrections
   to mass-dependent observables},
    Phys. Lett. B402 (1997) 359--366.
    \newblock
     \href{http://arxiv.org/abs/hep-ph/9703226}
          {\path{arXiv:hep-ph/9703226}}.

\bibitem{PeSt02}
 A.~A. Penin, M.~Steinhauser,
  {Heavy Quarkonium Spectrum at $O(\alpha_s^5m_q)$ and Bottom/Top Quark Mass Determination},
   Phys. Lett. B538 (2002) 335--345.
   \newblock
    \href{http://arxiv.org/abs/hep-ph/0204290}
               {\path{arXiv:hep-ph/0204290}},
    \href{http://dx.doi.org/10.1016/S0370-2693(02)02040-3}
               {\path{doi:10.1016/S0370-2693(02)02040-3}}.

\bibitem{KuSt01}
 J.~H. K{\"u}hn, M.~Steinhauser,
  {Determination of $\alpha_s$ and heavy quark masses from recent measurements of $R(s)$},
   Nucl. Phys. B619 (2001) 588--602.
   \newblock
    \href{http://arxiv.org/abs/hep-ph/0109084}
          {\path{arXiv:hep-ph/0109084}},
    \href{http://dx.doi.org/10.1016/S0550-3213(01)00499-0}
          {\path{doi:10.1016/S0550-3213(01)00499-0}}.

\bibitem{BM08}
 A.~P. Bakulev, S.~V. Mikhailov,
  {Resummation in (F)APT},
   in: A.~P. Bakulev \textit{et al.} (Eds.),
   Proceedings of International Seminar on Contemporary Problems of Elementary
   Particle Physics, Dedicated to the Memory of I.~L.~Solovtsov,
   Dubna, January 17--18, 2008., JINR, Dubna, 2008, pp. 119--133.
   \newblock
    \href{http://arxiv.org/abs/0803.3013}
         {\path{arXiv:0803.3013 [hep-ph]}}.

\bibitem{KK09}
 A.~L. Kataev, V.~T. Kim,
  {Uncertainties of QCD predictions for Higgs boson decay into bottom quarks at NNLO and beyond},
   PoS ACAT08 (2009) 004.
   \newblock
    \href{http://arxiv.org/abs/0902.1442}
         {\path{arXiv:0902.1442}}.

\end{thebibliography}

\newcommand{\noopsort}[1]{} \newcommand{\printfirst}[2]{#1}
 \newcommand{\singleletter}[1]{#1} \newcommand{\switchargs}[2]{#2#1}

\end{document}